\newif\ifonecol 
\onecolfalse 

\ifonecol
\documentclass[journal,12pt,onecolumn]{IEEEtran}
\else
\documentclass[journal]{IEEEtran}
\fi

\newlength{\figurewidth}
\ifonecol
\else
\fi

\ifonecol
\usepackage{setspace}
\fi

\usepackage{amsmath}
\usepackage{amssymb}
\usepackage{graphicx}
\usepackage{caption}
\usepackage{subcaption}
\usepackage{colortbl}
\usepackage[hidelinks]{hyperref}
\usepackage{algpseudocode}

\usepackage{lineno}

\ifCLASSOPTIONcompsoc
  \usepackage[nocompress]{cite}
\else
  \usepackage{cite}
\fi

\ifCLASSINFOpdf
\else
\fi

\newcommand{\grad}{$^{\circ}$}

\usepackage{float} 

\newcommand{\MYfooter}{\smash{\scriptsize
\hfil\parbox[t][\height][t]{\textwidth}{\vspace*{-0.5cm}\centering
\copyright 2020 IEEE. Personal use of this material is permitted. Permission from IEEE must be obtained for all other uses, including reprinting/republishing this material for advertising or promotional purposes, collecting new collected works for resale or redistribution to servers or lists, or reuse of any copyrighted component of this work in other works. DOI: 10.1109/TIM.2020.3034986}\hfil\hbox{}}}

\makeatletter

\def\ps@IEEEtitlepagestyle{%
\def\@oddfoot{\MYfooter}%
\def\@evenfoot{\MYfooter}}

\makeatother

\begin{document}
\bstctlcite{IEEEexample:BSTcontrol}
\title{Denoising Atmospheric Temperature Measurements Taken by the Mars Science Laboratory on the Martian Surface}
\author{S. Zurita-Zurita, Francisco J. Escribano,~\IEEEmembership{Senior Member,~IEEE,} J. S\'aez-Landete and J.A. Rodr\'iguez-Manfredi
\thanks{S. Zurita-Zurita and J.A Rodr\'iguez-Manfredi are with the Astrobiology Center (INTA-CSIC), Madrid, Spain (email: szurita@cab.inta-csic.es, manfredi@cab.inta-csic.es).}
\thanks{Francisco J. Escribano and J. S\'aez-Landete are with the Signal Theory and Communications Department, University of Alcal\'a, Madrid, Spain (email: francisco.escribano@ieee.org, jose.saez@uah.es).}
}

\IEEEtitleabstractindextext{%
\begin{abstract}
In the present article we analyze data from two temperature sensors of the  Mars Science Laboratory, which has been active in Mars since August 2012. Temperature measurements received from the rover are noisy and must be processed and validated before being delivered to the scientific community. Currently, a simple Moving Average (MA) filter is used to perform signal denoising. The application of this basic method relies on the assumption that the noise is stationary and statistically independent from the underlying structure of the signal, an arguable assumption in this kind of harsh environment. In this paper, we analyze the application of two alternative methods to process the temperature sensor measurements: the Discrete Wavelet Transform (DWT) and the Hilbert-Huang Transform (HHT). We consider two different datasets, one belonging to the current Martian measurement campaigns, and the other to the Thermal Vacuum Tests. The processing of these datasets allows to separate the random noise from the interference created by other systems. The experiments show that the MA filter may provide useful results under given circumstances. However, the proposed methods allow a better fitting for all the realistic scenarios, while providing the possibility to identify and analyze other interesting signal features and artifacts that could be later studied and classified. The large amount of data to be processed makes computational efficiency an important requirement in this mission. Considering the computational cost and the filtering performance, we propose the method based on DWT as more suitable for this application.
\end{abstract}

}

\maketitle
\IEEEdisplaynontitleabstractindextext
\IEEEpeerreviewmaketitle

\ifCLASSOPTIONcompsoc
\IEEEraisesectionheading{\section{Introduction}\label{sec:introduction}}
\else
\section{Introduction}
\label{sec:introduction}
\fi

%
%
%
%

\IEEEPARstart {F}{rom} the decade of 1960, numerous space missions have been trying to reach the Red Planet in order to collect information and study its atmosphere and soil, and reveal whether there were past or present signs of life \cite{viking,pathfinder}. The Mars Rover Curiosity (part of the Mars Science Laboratory) is one of the missions that is currently on the Martian surface, and it has been sending very valuable information about the planet since 2012, from its landing location and immediate surroundings, at the Gale crater \cite{curiosity}. 

Among other sensors and instruments, the Curiosity rover is equipped with a meteorological station, better known as REMS (Rover Environmental Monitoring Station) \cite{rems}. Since its landing time, the scientific community has been making use of data from REMS to perform intensive research, such as studies about the formation of frost \cite{frost}, about the Martian surface temperature \cite{2015SurfaceTempData}, or to test diverse hypotheses about the Martian atmosphere \cite{rems100soles,rems269soles}. The cited articles are just some examples of the quantity of publications where REMS data have been exploited. We want to stress the fact that this is the first time a rover has provided measurements from Martian surface for such a wide period of time (with just scarce gaps in between), and with such temporal resolution. Therefore, their processing through different methods for the first time is endowed with intrinsic novelty. For this reason, due to their relevance, it is extremely important to present reliable REMS measurement results.

This article is focused on the study of data coming from the two Air Temperature Sensors (ATS) which are part of REMS \cite{ATS,ATS2}. The ATS sensors are responsible for providing a reliable temperature estimation of the local environment. They are exposed to an amount of thermal influences, such as the heat conduction through the rover body, and to other internal and external perturbations, such as the wind, the solar radiation or the electronic noise of the sensors. All this may make it difficult to provide an accurate temperature estimation without an appropriate processing of the raw data. Moreover, the novelty and specificity of the REMS sensor suite and instrumentation, along the uncertainties still to be addressed by the science respecting the Martian atmospheric conditions and dynamics, determines the lack of an appropriate previous background to resort to. Therefore, this initial approach for processing REMS temperature measurements should reasonably rely on standard denoising methods, adapted to this specific context.

There are many denoising methods already present in the literature, useful for a wide range of applications. Classical methods, based on Fourier analysis, are widely used in filtering applications. However, as they lead to the loss of temporal information from the signals, there is an intrinsic limitation to their performance in extracting information from non-stationary processes. In addition, they mask and/or distort certain artifacts present in the signal \cite{proakis2013digital}. To avoid the loss of temporal information, the Short Time Fourier Transform and spectrograms have been widely used. These methods allow a simultaneous time and frequency analysis, and the generalization and formalization of this concept have given rise to different families of wavelet transforms \cite{mallat1999wavelet}. Other methods, like Independent Component Analysis (ICA), Sparse Spectral Decomposition and Reconstruction, and Singular Spectrum Analysis are also widely known alternatives \cite{ghilSSA}, but they assume some hard statistical hypothesis which are not fully fulfilled in our target data, e.g. that the different components are statistically independent, that they are uncorrelated, or that they are wide sense stationary processes.

A different and interesting non-linear approach to the same problem relies on the Hilbert-Huang Transform (HHT), which performs a decomposition of the signal into oscillatory modes with variable amplitude and instantaneous frequency. This variability, as opposed to the fixed amplitude and frequency of the harmonic decomposition of Fourier analysis, generates an additional degree of flexibility to process more challenging signals \cite{huang2014hilbert}.

Besides the HHT, wavelets are also widely known for their applications and have been used in many different areas in signal denoising and feature detection. Wavelets have advantages over traditional transforms when representing functions that have discontinuities and sharp peaks, and for accurately deconstructing and reconstructing finite, non-periodic and non-stationary instrumentation signals \cite{waveieee4,waveieee3,waveieee2,waveieee,waveletRaman,Speechwavelet,calibrationwavelet,EGCDscontinuitieswavelet,Armonicswavelet}. The HHT, on the other hand, has been designed to offer a high degree of flexibility and adaptability for the processing of signals with random components (e.g. stochastic data plus noise) \cite{emdieee,emdhhtieee,EMDHuang}, which is the case of the registered ATS measurements. According to this, both methods may offer added valuable possibilities for the processing of the REMS ATS data, as they can help in extracting interesting signal features, apart from denoising.

Currently, to mitigate the disturbances in the ATS data, specifically in the form of noise, a moving average (MA) filter is being used. This process often masks certain signal artifacts, and this may make further analysis difficult. The main purpose of the present article is to introduce the Discrete Wavelet Transform (DWT) and the HHT as more appropriate methods to denoise the REMS ATS measurements, while preserving and allowing the location of other signal features for further research and analysis. This is even more critical if we recall that, as studied and characterized in the related literature (e.g. \cite{ghilSSA}), detected abrupt shifts in processed environmental data may help in revealing relevant climatic features. Another example why keeping signal integrity is a key factor is shown in a recent published work \cite{SORIASALINAS2020113785}, where data from the ATS are used to provide extra information to support REMS Wind Sensor (WS) readings. As it is currently not working, the authors propose an indirect method to provide wind speed information from the processing of the ATS data. This shows why it is advisable to be conservative in the denoising process, so as not to excessively smooth the ATS measurements.

As part of our work, we have also determined which of the main mentioned methods are better fit to denoise the signal, while keeping as much information from artifacts and other relevant features for subsequent study.

It is to be noted that the scientific community using data from similar missions might take inspiration from the proposed denoising methods, as the instruments involved may face similar issues.
Given the previously described context, the main contributions of this article can be summarized as follows:
\begin{itemize}
 \item We have adapted the DWT and HHT methods to the specific context of MSL recorded Martian air temperatures.
 \item We have determined which method can be more suitable from the point of view of distortion, signal feature preservation and processing time.
 \item We have cast the grounds to extend the results of the present study to the identification and characterization of relevant perturbing phenomena in the corresponding measurements.
\end{itemize}

According to these aims, the article is structured as follows: Section \ref{data_processing} describes the instrument itself, the current data processing flow and its most relevant issues. Section \ref{method_analysis} provides a description of the DWT and the HHT as alternatives to offer better temperature estimations, and explains how each method has been adapted to the REMS ATS measurements. Section \ref{results} explains the Thermal Vacuum Test (TVT), how it can be used as a reference, and shows comparative results using the different alternatives, ascertaining their appropriateness. Section \ref{conclusions} is devoted to the conclusions. 
\begin{figure}[htbp]
	\centering	
	\includegraphics[width=8.8cm]{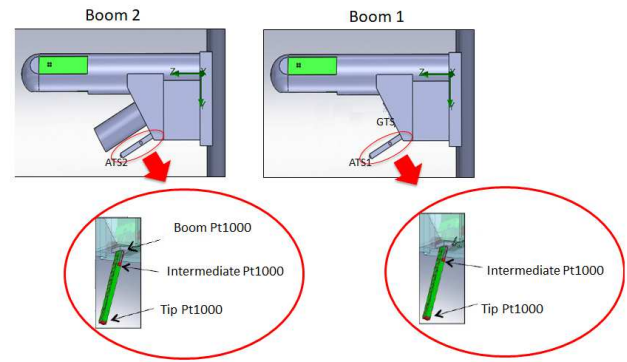}
	\captionsetup{font=scriptsize}
	\caption{Detailed depiction of part of the ATS instrument for each boom.}
	\label{fig:ats1boom}
\end{figure}
\begin{figure}[htbp]
	\centering	
	\includegraphics[width=8.8cm]{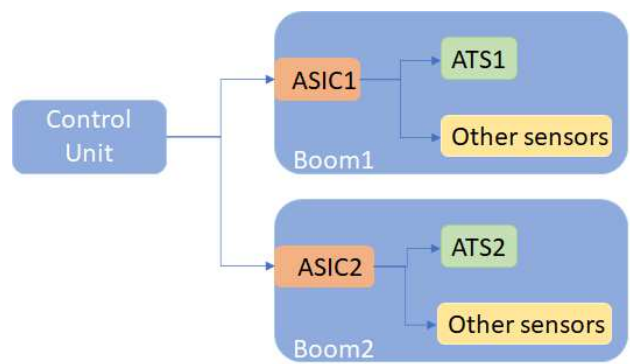}
        \captionsetup{font=scriptsize}
	\caption{REMS connection schematics.}
	\label{fig:sensors}
\end{figure}

\section{Current data processing flow in REMS ATS}\label{data_processing}
REMS is composed of a suite of sensors, and many of them are located within two booms\footnote{Due to their protuberant form, each of the bodies containing the majority of the REMS sensors was denominated as `boom'. }. The booms are placed perpendicularly to the rover main mast \cite{rems}. Each boom contains an Application-Specific Integrated Circuit (ASIC), whose function is to control and manage the different sensors, including the ATS sensors (ATS1 and ATS2). The ATS1, mounted on the so-called Boom 1, is equipped with 2 Pt1000 sensors, glued to the tip and to an intermediate position of a small FR4 rod, respectively. The Ground Temperature Sensor (GTS) is placed at the base of the ATS1, which, among other functions, provides the temperature at this point and takes the place of the Pt1000 at the base of the FR4 rod of the Boom 2.  The ATS2, placed on the so-called Boom 2, has 3 Pt1000 sensors, glued to the tip, to an intermediate position and to the base of a similar FR4 rod, respectively. Both booms can be seen in Figure \ref{fig:ats1boom}. Figure \ref{fig:sensors} shows the basic schematics of the main REMS subsystems and their connections within each boom. We can see how they are both linked to their respective ASIC. The ATS sensors work at a sampling frequency of $1$ sample per second.

Both ATS sensors are constantly exposed to a number of external and internal perturbations. Externally, each ATS is exposed to solar radiation and potentially strong Martian winds. Internally, the ATS's are mainly affected by the thermal contamination generated by the rover power supply, a Radioisotope Thermoelectric Generator (RTG), which is the warmest location in the rover Curiosity. Other meaningful internal perturbations are related to the switching on and off of other sensors belonging to REMS, such as the WS, or the changes in configuration of the ASIC itself, which is shared with other sensors and may introduce disturbing effects to the temperature readings. Due to all this, it is difficult to provide a reliable temperature measurement without further processing.

Currently, the ATS REMS data downloaded to Earth are processed as follows:
\begin{enumerate}
\item The ATS REMS data reach the Earth through eight data channels. Five channels contain information from the five Pt1000 sensors from both ATSs, and the other three channels contain information about the temperature readings from the GTS, corresponding to three different infrared wavelength channels. At this stage, the data are in a raw format, better known as $counts$, which represents the readings related to the physical magnitude corresponding to each physical sensor.
\item The counts are translated into kelvin (K) by using the Pt1000 physical model and its conversion equations \cite{pt1000conversion}. This gives the temperature measured by the sensors, and will serve as the basis for the estimation of the air temperature. The physical model takes into account certain corrections depending on other sensors status, such as the gain of the GTS or the activation/deactivation state of the WS. The model considers six measurement signals, one from each of the five Pt1000 of both booms, while the temperature at the ATS1 base is obtained from the three GTS readings, as mentioned before. The signals obtained from each sensor, already translated to kelvin, will be denoted from now on as $x_k[n]$, where the indexes $k=1,2,3$ identify the three signals from the temperature sensors of Boom $1$ (a Pt1000 at the tip, a Pt1000 at an intermediate point and the mean value of the three thermopiles  of the GTS located at the boom base, respectively), and the indexes $k=4,5,6$ identify the signals from the corresponding Pt1000 sensors of the Boom $2$ (tip, intermediate point and base, respectively). The signals are considered after the analog-to-digital conversion (ADC) step, hence the discrete-time nature of the defined variables.
\item\label{item3} A filter is applied to each of the signals in order to reduce the noise, so that we can write
\begin{equation} y_k[n] = x_k[n] \ast h[n], \end{equation}
where $y_k[n]$ is the filtered version of the $k$-th signal, $h[n]$ is the impulse response of the filter and $\ast$ denotes the convolution operator. As stated, an MA filter is the method currently applied for this task. Its application is based on the assumption that the noise is statistically independent from the temperature evolution and, therefore, it should not change the underlying structure of the signal. Under this hypothesis, averaging over a few points would effectively reduce the noise contribution. Specifically, for each data sample, it calculates the average over a predefined number of neighbors, defined by a span parameter. As it is widely known, this is an instance of a Finite Impulse Response (FIR) lowpass filter, where the span is related to the cutoff frequency. The computational cost is low but the stopband attenuation, approximately $21$ dB, is often insufficient \cite{proakis2013digital}. Increasing the span will reduce the cutoff frequency and the effects of the noise, at the expense of a noticeable signal distortion.

The MA filter, in fact, provides useful results for signals that are lowpass, continuous and smooth. However, when abrupt or fast changes are present --and they may arise from artifacts that may contain meaningful information-- the filtering process might mask or distort them. Accordingly, this filter cannot be used if we want to get further insights about complex phenomena affecting the measurements.  The calibration of the ATS sensors and the experimental sensor response time were dominant factors to determine the span value best fitted to the nature of the ATS signals. The criterion chosen to set the MA filter was extremely conservative to avoid removing at least certain relevant artifacts, and it was decided that a reasonable trade-off for the span would be $9$ samples. The coefficients used for the MA filter are constant, and the impulse response $h\left[n\right]$ is defined as follows:
\begin{equation}h[n] = \frac{1}{9}\sum_{k=0}^{8}{\delta[n-k]}, \end{equation}
\item The retrieval model, whose details can be found in \cite{ATS}, extrapolates the ambient temperature using the previously processed temperature data from the three sensors at each ATS. The thermal model of the FR4 beam is based on the theory of heat transfer from a constant section fin surrounded by a fluid. In this context, a fin is a surface protruding out from an object, whose function is to increase the heat transfer rate to or from the environment by offering a maximal exposure surface area. The FR4 beam can be modeled as such a fin, transferring heat from the boom to the environment. This model, based on the principle of an infinite fin \cite{mueller}, is applied to give a local temperature estimation next to the tip by using all the temperature measurements from both ATSs. The estimated ambient temperature of each boom will be denoted as
\begin{align}
t_1[n] & = \text{R}\{y_1[n], \; y_2[n], \; y_3[n]\}, \nonumber \\ 
t_2[n] & = \text{R}\{y_4[n], \; y_5[n], \; y_6[n]\},
\end{align}
where the application of the retrieval method is represented by the operator $\text{R}\{\}$, and $t_1[n]$ and $t_2[n]$ are the estimated ambient temperatures from the Boom $1$ and the Boom $2$, respectively.
\item At the end of the processing chain, a unique ambient temperature estimation $t[n]$ is obtained from the ATS1 and ATS2 estimated temperatures. Due to the low temperatures measured on Mars during the whole Curiosity mission, it is considered that the ambient temperature closest to the real value should be the coolest one among $t_1[n]$ and $t_2[n]$, specially because the rover (and its RTG) may significantly contribute to an undesirable increase in the temperature of the air surrounding it.
This can be denoted as
\begin{equation} \label{eqn:1} t[n] = \min( t_1[n], \; t_2[n]). \end{equation}
\end{enumerate}

We consider that it is possible to improve the ATS data processing, most specifically with respect to the filtering process described in step \ref{item3}, as it will be made more evident in the sequel. We propose here the usage of alternative methods which may fit better the structure and dynamics of the observed signals, because the moving average filter cannot work efficiently when the signals are not stationary and lowpass. We will focus on the DWT and the HHT, which have a solid background in similar applications, and will be specifically adapted to the phenomenology of the ATS instrument, as detailed in the following section.

\section{Analysis and application of alternative signal processing methods}\label{method_analysis}
As explained in the previous section, the MA filter gives as result six different filtered signals, denoted as $y_k[n]$. The limitations of this method and its lack of further in-depth analysis capabilities about the results pose certain concerns about its suitability. For this reason, more elaborated methods may be considered in order to improve the quality of the data presented to the scientific community.
 After having performed a thorough research, and knowing the specificity of our data, we have chosen two possible algorithms. One of the possibilities are the wavelets (and specifically the DWT), a family of functions that has proven to be specially useful in signal denoising. A second possibility is the HHT algorithm, whose main feature is its high versatility for the processing of random data. In order to verify the suitability of these methods, they have been applied to two different kinds of datasets: on the one side, measurement data recorded under real conditions on Martian surface, and, on the other side, a set of test measurements theoretically unaffected by external perturbations.

In order to simplify the notation and without loss of generality, we will describe the denoising procedure based on each of the above mentioned methods for a generic signal $x[n]$, where we have dropped the subindex for convenience. It is to be understood that these methods will be independently applied to each of the six original signals, $x_k[n],\; k=1,\ldots 6$.

\subsection{Denoising with the DWT}
The data processing based on the DWT is detailed as follows:
\begin{enumerate}
\item The DWT is applied to the ATS data.

Wavelets analysis makes use of short-duration waveforms, $\psi[n]$, with zero mean and a sharp decay to zero at both ends. These short-duration waveforms are scaled and shifted to set the optimum time-frequency resolution. We use the definition and formulation of Mallat \cite{mallat1999wavelet}, where the discrete wavelet scaled by $a^j$ is expressed as
\begin{equation} \label{eqn:scaled_wavelet}  \psi_j[n] = \frac{1}{\sqrt{a^j}}  \psi\left[\frac{n}{a^j}\right] \end{equation} 
where $j$ is the subband or scale level of the decomposition. The scale factor chosen is $a=2$ since dyadic decomposition is used. The DWT coefficients of the signal $x[n]$ with respect to the wavelet function $\psi[n]$ can be written as
\begin{equation} \label{eqn:wave}  X_j[m] = \sum_{n=0}^{N-1}x[n]\cdot \psi_j^*\left[n-m\right], \end{equation}
where $\psi_j^*\left[n\right]$ is the complex conjugate of the scaled wavelet, $m$ is the translation parameter, and $N$ the number of samples of the noisy signal. In similar applications \cite{waveletSelection}, the choice of the mother wavelet is mainly based on visual inspection and on the correlation between the original signal and the wavelet-denoised signal. The wavelets families Daubechies, Symlet and Coiflets have been used to test which one may be better fitted to the nature of our data. After a comparative analysis, we concluded that the Coiflet 5 was the one determining better reconstruction results for this specific application.

It is worth noting that equation \eqref{eqn:wave} can be expressed as a circular convolution, so that we can resort to the Fast Fourier Transform when making the corresponding computation. This approach just requires $O(N\log_2 N)$ operations for each scale, resulting in high computational efficiency. This feature can be seen as a significant advantage over the HHT, whose complexity is well over the previously mentioned result.
 
\item Thresholding of the subband signals with a threshold $\theta$.

After having decomposed the original signal, a thresholding step is applied to the coefficients of the wavelet decomposition. Its purpose is finding out which part of them should be qualified as noise, and should be consequently removed. For this purpose, the universal threshold from Donoho and Johnstone has been used \cite{donoho1993Johnstone}. The threshold is calculated using the coefficients from the finest decomposition level (in our case $j=1$). They are related to the higher signal frequencies, where the main contribution from the noise is supposed to be found. This unique threshold is used for all the coefficients of the different scales. Said threshold is calculated as \cite{donoho1993Johnstone}
\begin{equation}\label{eq:threshold} 
\theta = \sigma \cdot  \sqrt{2\cdot \log (N)},
\end{equation}
where $N$ is the length of the signal, and $\sigma$ is the estimated standard deviation of the noise, which is calculated using the so-called Median Absolute Deviation (MAD) from the finest decomposition level of the wavelet transform, $X_1[m]$, as \cite{donoho1993Johnstone}
\begin{equation}\label{eq:sigma}
\sigma = \frac{\text{MAD}}{0.6745}.
\end{equation}
The MAD is defined as
\begin{eqnarray}  \text{MAD} = \text{med}\left\{\left| X_1[m] -  \text{med}\left\{X_1[m]\right\}\right|\right\},\end{eqnarray}
where $\text{med}\{\}$ is the median value. These expressions take into account that the value of the noise standard deviation $\sigma$ is often estimated from the median value of the $X_j[m]$ wavelet coefficients belonging to the first level of signal decomposition \cite{Somekawa2013}. 

There is the possibility to perform hard or soft thresholding, but, after having done some tests with both alternatives, we have concluded that soft thresholding actually gives better results for our specific signals. According to this, the thresholding process is applied, as detailed in \cite{donoho1995}, for each zero-mean decomposition level, in order to obtain the thresholded coefficients $\hat{X}_j\left[m\right]$ as

\begin{align}
\hat{X}_j[m] = 
\begin{cases}
X_j[m] - \theta,  &  X_j[m] \ge  \theta \\
0, &  |X_j[m]|<  \theta \\
X_j[m] + \theta, &  X_j[m] \le -  \theta.
\end{cases}
\end{align}
Following \cite{donoho1993Johnstone}, Section 2.4, the wavelet levels that do not have vanishing means should not be thresholded, since this would lead to severe distortion.
\item Reconstruction of the filtered output signal.

As described, once the thresholding has been applied to $X_j[m]$, the output coefficients $\hat{X}_j[m]$ are obtained. They are used to produce the reconstructed signal $y^{w}[n]$ by means of the inverse DWT,
\begin{equation}\label{eqn:waveoutput}  
y^{w}[n] = \sum_{m=0}^{N-1}\sum_{j=1}^{J}{ \hat{X}_j[m]\cdot \psi_j\left[m-n\right]},\end{equation}
where $J$ is the number of decomposition levels. In our developments, the maximum decomposition level permitted has been used, according to the length of the signal $x[n]$.
\item By subtracting the reconstructed signal from the original one $x[n]$, we obtain an estimation of the perturbing noise $\epsilon^{w}\left[n\right]$ as
\begin{equation} 
\epsilon^{w}[n] = x[n] -  y^{w}[n].
\end{equation}
\end{enumerate}

\subsection {Denoising with the HHT}
The denoising based on the HHT method is detailed in the sequel, according to \cite{boudraa2007empirical}. The exact procedure has been the result of a number of tests, where different number of Intrinsic Mode Functions (IMFs) and other parameters have been tuned in order to reconstruct the signal, according to the nature of our data. The method proceeds as follows:
\begin{enumerate}
\item The HHT is applied to the ATS data.  

The fundamental part of the HHT is the Empirical Mode Decomposition (EMD), a fully data-driven approach. Using this method, a noisy signal is adaptively decomposed into oscillatory modes with variable amplitudes, called IMFs, plus a residual signal. An IMF is a function where the number of extrema and zero-crossings must differ at most by one, and the mean of the envelope defined by linking the local maxima and local minima is zero. The procedure of extracting an IMF is called {\it sifting}, and it consists on the subtraction of the mean of the envelope defined by the local maximum and local minimum from the signal. If the subtraction is not an IMF, the difference is considered a new signal and should be iteratively subtracted. When an IMF is found, the difference between the original signal and the IMF becomes the new signal, and a subsequent sifting process is performed to find the next IMF. The thresholds used for the stopping criterion were chosen as $0.05$ and $0.5$. The decomposition of the signal in $M-$empirical modes can be written as
\begin{equation}\label{eqn:EMD}  {y}[n] = \sum_{m=1}^{M} c_{m}[n] + r[n] , \end{equation}
where $c_{m}\left[n\right]$ is the IMF of the $m$-th decomposition level, and $r[n]$ is the residual. This latter is a monotonic function from which no more IMFs can be extracted. The signal ${y[n]}$ is the reconstruction after applying the EMD to the input data $x[n]$. Under ideal conditions, without further processing of the IMF values, the reconstructed signal should be equal to the input signal, i.e. ${y[n]}=x[n]$.  

\item Apply the Consecutive Mean Square Error (CSME).

This step is mainly based on the idea that the main part of the meaningful signal structure has to be found in the last IMFs, which represent the lower frequencies, while the noise is usually associated to the first IMFs, which represent the higher frequencies. Accordingly, the denoising process consists on the reconstruction of the signal after discarding the initial IMF levels. Therefore,
\begin{equation}
{y_k}[n] = \sum_{m=k}^{M} c_{m}[n] + r[n]
\end{equation}
is the reconstructed signal taking into account just the last $M-k+1$ levels. We consider the CSME as a measure of the distortion in the reconstructed signal, according to \cite{boudraa2007empirical}. The CMSE measures the squared Euclidean distance between two consecutive reconstructions of the signal, and it is defined as
\begin{align}  
\text{CSME}(y_{k},y_{k+1}) & = \frac{1}{N} \sum_{n=0}^{N-1} |y_k[n]-y_{k+1}[n]|^2 =\nonumber \\
& = \frac{1}{N} \sum_{n=0}^{N-1} |c_{k}[n]|^2.  
\end{align}
The index $j$ that minimizes the CSME allows to determine which $\text{IMF}$ level represents the limit between the part of the signal where the noise can be considered negligible, and the part where the noise is dominant \cite{boudraa2007empirical}. The index $j$ is given by calculating
\begin{equation}
j = \underset{1\leq k \leq M-1}{\mathrm{arg min}}(\mathrm{CMSE}(y_{k}, y_{k+1})).
\end{equation}
Therefore, the signal reconstructed using the decomposition levels from this index and on should mainly contain the noise-free signal components. A large number of experiments have been performed and reported in the literature to support this conclusion \cite{boudraa2007empirical}. The IMFs from index $j+1$ to the last one (plus the residual) should be related to the structure of the noise-free signal. However, in our case, the first IMFs could contain meaningful signal artifacts, and therefore we consider the inclusion of the first IMFs processed through thresholding.

\item Apply a threshold to the lowest IMF levels.

 As the noise is assumed to be distributed among all of the mentioned IMF levels ($\mathrm{IMF}_{0,..,.j}$), it is reasonable that a different threshold should be calculated for each of them \cite{boudraa2005}. In order to explicitly filter the additive Gaussian noise, the universal threshold applied to the wavelets is applied here, but it has to take different values for each decomposition level. If we denote the threshold for the $m$-th level as $\theta_{m}$, where $m=0,\cdots,j$,
\begin{equation}
\theta_{m} = \sigma_{m}  \cdot  \sqrt{2\cdot \log (N)},
\end{equation} 
where $N$ is the length of the noisy signal, and $\sigma_m$ the noise standard deviation of the $m$-th IMF. The estimated value for $\sigma_m$ is \cite{boudraa2006}
\begin{equation}
\sigma_{m} =\frac{\text{MAD}_{m}}{0.6745},
\end{equation}
where $\text{MAD}_m$ is the absolute median deviation of the $m$-th IMF, calculated as
\begin{equation}
\text{MAD}_m = \text{med}\left\{ \left| c_m[n] -  \text{med}\left\{ c_m[n] \right\} \right| \right\}.
\end{equation}
Again, a soft thresholding method is proposed to reconstruct the signal, and it is applied as \cite{donoho1995}

\begin{eqnarray}
\hat{c}_{m}[n] =
\begin{cases}
c_{m}[n] - \theta_{m},  &  c_{m}[n] \ge \theta_{m} \\
0, & |c_{n}[n]| < \theta_{m} \\
c_{m}[n] + \theta_{m}, & c_{m}[n] \le - \theta_{m}.
\end{cases}
\end{eqnarray}
This procedure is practically the same as the one proposed for the application of the wavelets, but in this case the threshold depends on the decomposition level, and only the first $j$ modes are processed.
\item Reconstruction of the filtered output signal.

After applying the thresholding to the corresponding levels, the reconstructed signal $y^{h}\left[n\right]$ is obtained by adding the thresholded IMFs, the IMFs from the $j+1$-th level and the residual, so that
\begin{equation}\label{eqn:EMDoutput} y^{h}[n] = \sum_{m=0}^{j}\hat{c}_{m}[n] +  \sum_{m=j+1}^{M}c_{m}[n] + r[n]. \end{equation}
\item By subtracting the reconstructed signal from the original one, $x[n]$, we obtain an estimation of the perturbing noise $\epsilon^{h}\left[n\right]$ as
\begin{equation}\epsilon^{h}[n] =  x[n] - y^{h}[n].\end{equation}
\end{enumerate}

\section{Results}\label{results}
In this section, we present the results from the application of the filtering methods described, as compared to the application of the moving average filter. Throughout this Section, all the results have been obtained by applying the afore-described algorithms codified in python, resorting to standard libraries like pywt for DWT, and pyhht for HHT. Before unfolding said results, we detail two additional sources of temperature measurement datasets that have been used to contrast and assess the goodness of the proposed approaches.

\subsection{Thermal Vacuum Test (TVT) and Cruise Checkout}
There are two sources of data that, in principle, can be considered practically unaffected by the typical external perturbations:
\begin{enumerate}
\item The so-called ATS Cruise Checkout data, a vacuum health status check performed during the cruise phase to Mars, where the ambient temperature was estimated to be $-10$ \grad C.
\item The so-called ATS Thermal Vacuum Test (TVT), which was performed before the mission launch. The test took place inside a vacuum chamber with a cooling screen placed at the boom base, and a surrounding cover refrigerated with liquid nitrogen. Under these conditions, intensive measurement tests were performed to check the electronic noise generated by the instrument at different target temperatures: $0$ \grad C, $-30$ \grad C, $-50$ \grad C, $-70$ \grad C and $-90$ \grad C.
\end{enumerate}

The noise affecting the datasets mentioned above, which are processed using the previously detailed methods, stems by hypothesis from the electronics surrounding the ATS. Under the described conditions, such data should be largely considered free from noticeable external perturbations.
%
\begin{figure}[htbp]
\centering
\includegraphics[width=8.8cm]{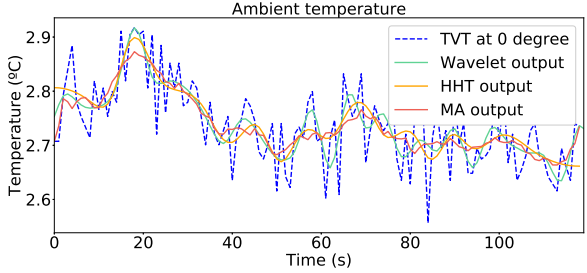}	
\captionsetup{font=scriptsize}
\caption{Results with a TVT dataset at $0$ \grad C. MA (span = 9), DWT (Coiflet 5) and HHT (stop criterion: 0.05 and 0.5). Notice that all of them yield similar results.}
\label{fig:Figu3}
\end{figure}

\subsection{Assessing the limitations of the moving average filter method}
In Figure \ref{fig:Figu3} we have depicted data from a nominal ATS TVT dataset, where there are no abrupt changes in the evolution of the average temperature. The blue dotted line, labeled ``TVT at 0 degree'', represents the original signal from the test at $0$\grad C, the line labeled ``Wavelet output'', in green, represents the filtered TVT signal after having applied the DWT based method of equations \eqref{eqn:wave} - \eqref{eqn:waveoutput}, the line labeled `` HHT output'', in orange, represents  the filtered TVT signal after having applied the HHT based method of equations \eqref{eqn:EMD} - \eqref{eqn:EMDoutput}, and the line labeled ``MA output'', in red, represents the filtered TVT signal after having applied the MA filter. We can thus visualize how each method performs the filtering of the noise. The MA filter is the method with the softest reconstruction because it makes use of a relatively large span, as mentioned earlier. As a consequence, in this situation, a loss of artifacts in the signal may easily occur. The HHT and the DWT based methods use more elaborated processing algorithms, as explained before, and are able to follow the original signal with greater accuracy by visual inspection. Nevertheless,  overall, the three methods do no offer much different features when processing a signal of this kind.

\begin{figure*}
\begin{subfigure}{.49\textwidth}
  \centering
  \includegraphics[width=.8\linewidth]{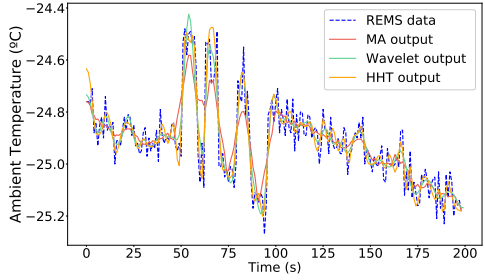}
  \captionsetup{font=small}
  \caption{Artifacts on Sol 120,  around 16:00.}
  \label{fig:sub-first}
\end{subfigure}
\begin{subfigure}{.49\textwidth}
  \centering
  \includegraphics[width=.8\linewidth]{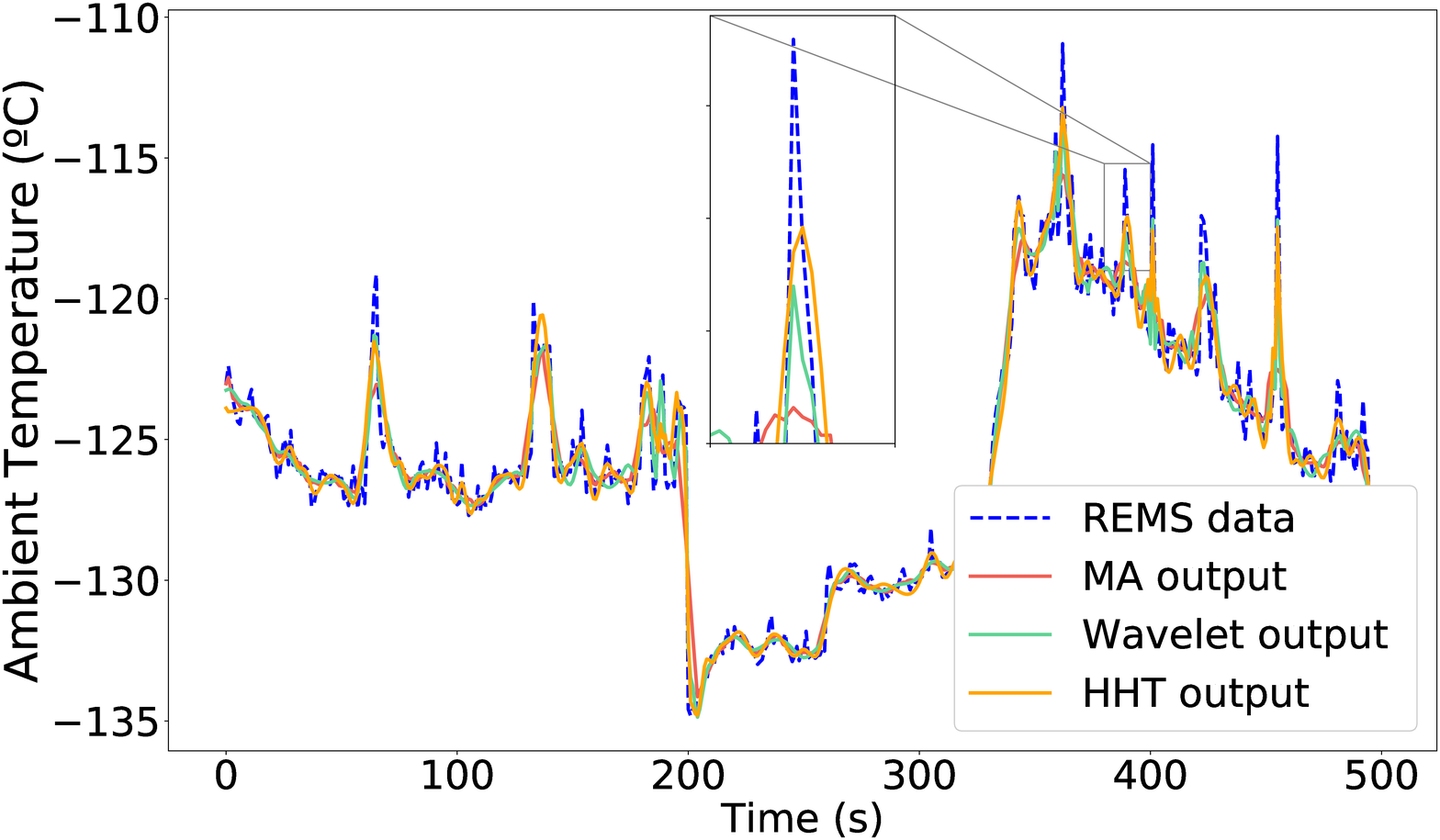}
  \captionsetup{font=small}
  \caption{Artifacts on Sol 68,  around 7:00}
  \label{fig:sub-second}
\end{subfigure}
\newline
\begin{subfigure}{.49\textwidth}
  \centering
  \includegraphics[width=.8\linewidth]{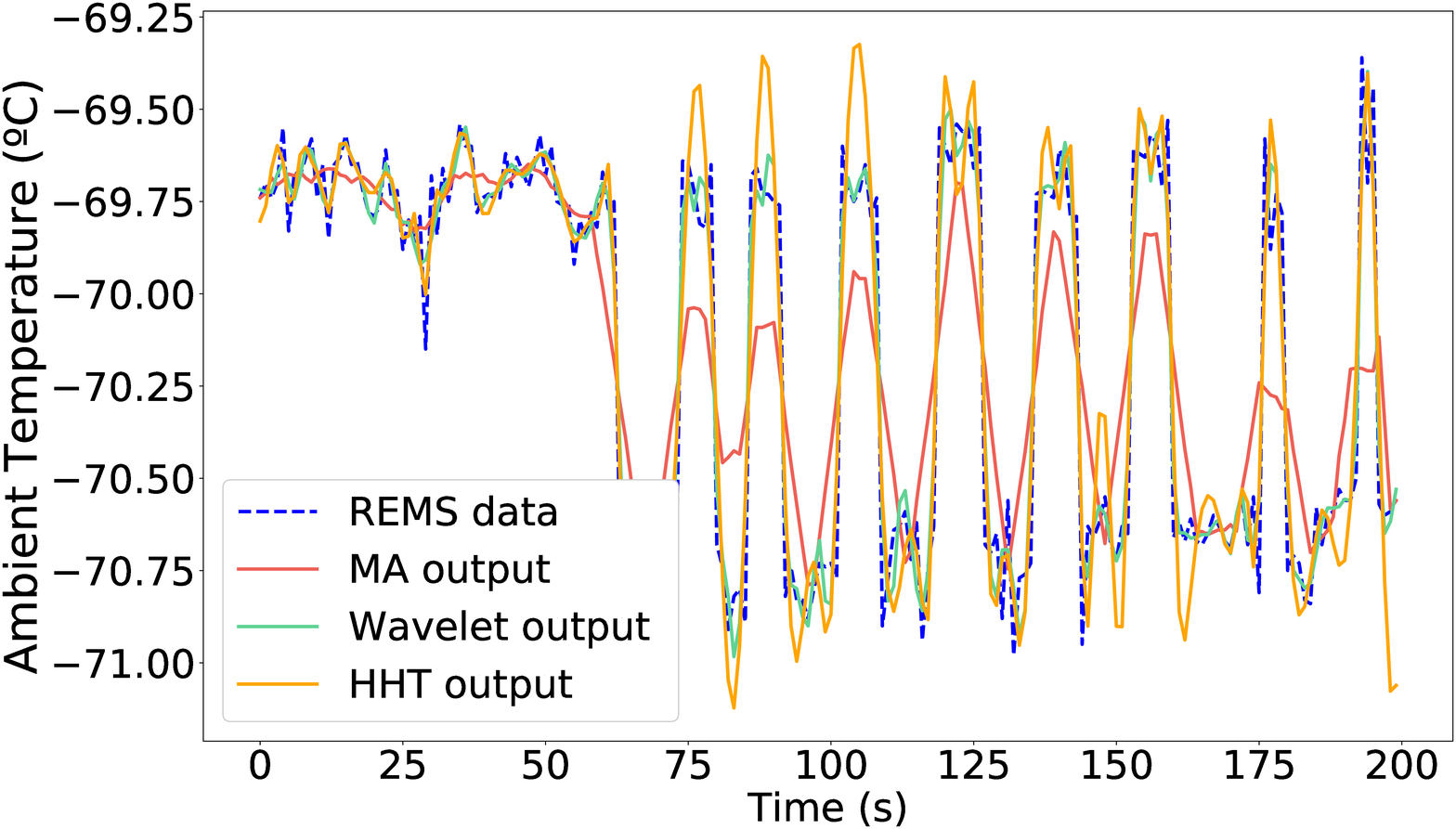}  
  \captionsetup{font=small}
  \caption{Artifacts on Sol 87,  around 4:00.}
  \label{fig:sub-third}
\end{subfigure}
\begin{subfigure}{.49\textwidth}
  \centering
  \includegraphics[width=.8\linewidth]{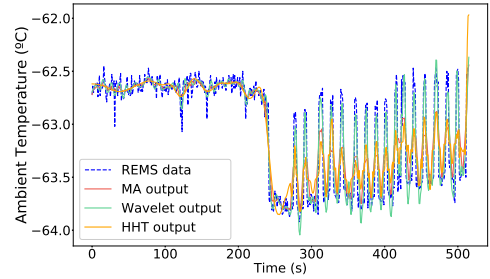}  
  \captionsetup{font=small}
  \caption{Artifacts on Sol 107, around 4:35.}
  \label{fig:sub-fourth}
\end{subfigure}
\captionsetup{font=scriptsize}
\caption{Set of significant Mars signals, with results from MA (span = 9), DWT (Coiflet 5) and HHT (stop criterion: 0.05 and 0.5) techniques. Notice that MA filtering excessively smoothes the signal.}
\label{fig:4figuras}
\end{figure*}

The MA filter limitations are made evident when the dataset contains abrupt changes in the measured temperatures. To confirm this, a characteristic dataset from REMS\footnote{Available at atmos.nmsu.edu/PDS/data/mslrem\_1001/DATA/.} \cite{solsData} has been used. The data belong to measurements in sols\footnote{Sol definition: Martian day.} $68$, $74$, $107$ and $120$, where we can see how some kind of sharp temperature variations are present. Though it is not within the scope of this work to define or classify these artifacts, we think it is not advisable to remove them in the denoising process. Due to the ATS sensitivity, we think that these signal features could provide extra information to derive other physical magnitudes and support other sensors, as evidenced in \cite{SORIASALINAS2020113785}. The plots in Figure \ref{fig:4figuras} show data from the previously mentioned sols. They make evident how each method behaves distinctively when processing the noise and reconstructing the filtered signal. The MA filter cannot follow the noisy signal as accurately as the HHT or DWT based filtering, possibly losing relevant features for the study of local atmospheric temperature variations or for the discrimination of signal disturbances. Therefore, we can say that the current processing method softens the abrupt signal changes excessively. The zoomed-in areas shown in the plots of Figure \ref{fig:4figuras} highlight how the MA filter fails to follow the signal evolution, in contrast with any of the other two methods detailed. For this reason, more powerful denoising algorithms such as the ones based on the HHT or the DWT should be used when abrupt changes may appear, as it is very often the case with the data taken in the Martian atmosphere.

\subsection{Denoising signals with MA, DWT and  HHT}

In order to analyze the performance of the proposed methods in separating the noise from the target signal, and to assess the corresponding filtering potential, we compare the results when using the ATS TVT with the results when using the ATS data taken on Mars. The TVT signals are supposed not to contain external disturbances, and may be used as a reference to evaluate the denoising performance in conditions without noticeable signal distortion. To perform fair comparisons, it has been necessary to select datasets from Mars measurement campaigns where we have similar temperatures and exactly the same sensor configuration. To obtain the estimated noise signal of both the ATS TVT data and the Martian ATS data, we have to apply the MA, the DWT and the HHT methods as explained before.

As previously seen, the noisy part of the signal is estimated by subtracting the reconstructed signal, using the DWT or the HHT, from the original one, $x[n]$. This can also be applied to the MA filter method. The standard deviation is used here as the comparison parameter, because it is directly related to the estimated noise power. If the standard deviation values are similar when using a controlled signal dataset (i.e. TVT data) or a very perturbated dataset from Mars, it may be inferred that the corresponding method is effective in filtering the noise without affecting relevant signal information, because we can assume that we are essentially removing the electronic noise contribution.

Table \ref{table:standardDeviationWSOFF} shows the estimated standard deviation of the noise when using the MA filter, the DWT and the HHT, in the case when we process ATS TVT data and when we process ATS data from Mars, at different nominal temperatures. First of all, we corroborate that the estimated noise variance is quite similar for any of the three methods. If this was the only figure-of-merit, any of the three methods would be equivalent. However, when a perturbation is present in the target signal, the estimated standard deviation of the noise is clearly higher in the case of the MA with respect to the values attained with HHT and DWT. Table \ref{table:standardDeviationPerturbations} shows the numerical results for the signals in Figure \ref{fig:4figuras}. The data in Table \ref{table:standardDeviationWSOFF} and the first three rows in Table \ref{table:standardDeviationPerturbations} can be easily compared for close nominal temperatures, so that we can see a high degree of compatibility between the standard deviation of the noise estimated with the TVT data (which is basically the same irrespective of the technique used), and the standard deviation obtained by the application of the HHT and DWT methods. Notice that, in the case of the fourth row, with a nominal temperature around $-125$\grad C degrees, there is no TVT data to compare with. For these reasons, we discard the MA filter method, as stated, because of its inability to preserve other possible relevant signal features.

\begin{table}[htbp]
\captionsetup{justification=centering, font=scriptsize, labelsep=newline}
\caption{\textsc{Estimated standard deviation with Martian ATS data against the estimated standard deviation with TVT data, when applying MA, DWT and HHT methods. WS is deactivated.}}
\begin{center}
\begin{tabular}{c l c c c }
\hline
  \parbox{4em}{\centering Nominal temp \grad C} & \parbox{4.3em}{\centering Data from} &  \parbox{4.4em}{\centering $\sigma$ noise \grad C (DWT)}& \parbox{4.3em}{\centering $\sigma$ noise \grad C (HHT)}& \parbox{4.3em}{\centering $\sigma$ noise \grad C (MA)} \\ \hline
$0$ & TVT&$0.04$&$0.05$&$0.05$\\ 
$0$ & ATS (Mars) &$0.06$&$0.04$&$0.07$\\ 
$-10$ & Cruise &$0.05$&$0.06$&$0.06$ \\ 
$-10$ &  ATS (Mars) &$0.07$&$0.05$&$0.07$ \\ 
$-30$ & TVT &$0.04$&$0.04$&$0.05$\\ 
$-30$ & ATS (Mars) &$0.08$&$0.05$&$0.08$\\ 
$-50$ & TVT &$0.04$&$0.04$&$0.04$\\ 
$-50$ & ATS (Mars) &$0.08$&$0.06$&$0.09$\\ 
$-70$ & TVT &$0.05$&$0.07$&$0.06$\\ 
$-70$ & ATS (Mars) &$0.09$&$0.05$&$0.07$\\ 
$-90$ & TVT &$0.07$&$0.08$&$0.08$\\
$-80*$ & ATS (Mars) &$0.09$&$0.05$&$0.07$\\ \hline
\end{tabular}\vspace{0.2cm}
\label{table:standardDeviationWSOFF}
{\footnotesize *Note: In the last row we have -80 \grad C instead of -90 \grad C, because it is the closest temperature found in Mars data with the WS deactivated.}
\end{center}
\end{table}

\begin{table}[htbp]
\begin{center}
\captionsetup{justification=centering , font=scriptsize, labelsep=newline}
\caption{\textsc{Estimated standard deviation of the noise from Martian ATS data with perturbations, when applying MA, DWT and HHT methods.}}

\begin{tabular}{l c c c c }
\hline
 \parbox{4.2em}{\centering Data from} &  \parbox{3.8em}{\centering $\sigma$ \grad C (DWT)}& \parbox{3.8em}{\centering $\sigma$ \grad C (HHT)}& \parbox{3.8em}{\centering   $\sigma$ \grad C (MA)} \\ \hline
Noise from Sol 120&$0.064$&$0.078$&$0.1251$\\ 
Noise from Sol 87&$0.0935$&$0.1760$&$0.2865$\\ 
Noise from Sol 107&$0.1094$&$0.1700$&$0.2713$\\
Noise from Sol 68&$0.9465$&$0.9802$&$1.3773$\\ \hline
\end{tabular}\vspace{0.2cm}
\label{table:standardDeviationPerturbations}
\end{center}
\end{table}

\begin{table}[htbp]
\begin{center}
\captionsetup{justification=centering, font=scriptsize, labelsep=newline}
\caption{\textsc{PRD metrics.}}

\begin{tabular}{l c c c c}
\hline
 \parbox{4.2em}{\centering Data from} &  \parbox{3.8em}{\centering  DWT}& \parbox{3.8em}{\centering  HHT}& \parbox{3.8em}{\centering  MA} \\ \hline
Sol 120&$0.2592$&0.2525  &$0.3551$\\ 
Sol 87&$0.1706$&$0.2510$&$0.4084$\\ 
Sol 107&$0.2201$&$0.2711$&$0.3756$\\ 
Sol 68&$0.5568$&$0.6306$&$0.9562$\\ \hline
\end{tabular}\vspace{0.2cm}
\label{table:metric}
\end{center}
\end{table}

\subsection{PRD metric}
In order to quantitatively assess the goodness of the proposed methods, we have included a metric, normally used in the denosing literature to measure the distortion introduced in the filtering of a signal, i.e. the percentage root mean square difference (PRD) \cite{Sharma2010ECGSD}, defined as
\begin{equation}
PRD =\sqrt{\frac{\sum_{n=0}^{N-1} (x\left[n\right]-\hat{y}\left[n\right])^2}{\sum_{n=0}^{N-1} \left(x\left[n\right]\right)^2}}\cdot100,
\end{equation}
where $x\left[n\right]$ is the original data, and $\hat{y}\left[n\right]$ is the reconstructed signal. A lower PRD represents a better reconstruction of the signal after denoising. As can be seen in Table \ref{table:metric}, the current moving average filtering significantly increases the PRD of the signal, while the proposed techniques improve the denoising with reduced levels of distortion, in agreement with what is seen in the plots of Figure \ref{fig:4figuras}.

\subsection{Processing time}
Once we have seen that the DWT- and the HHT-based methods proposed can be equally efficient in processing the noise while keeping characteristic signal features, an additional issue to be considered in order to choose the best denoising method is the processing time consumption. This is an important requirement for the processing of REMS ATS data due to the fact that reports of results from all REMS sensors must be delivered in a very short period of time after the arrival of the raw data to Earth. Table \ref{table:comp} shows the time taken by the application of the three methods\footnote{The MA filter result is included for comparison.} to a dataset of $20$ MBytes, using an Intel Core processor i7, python 3.7 software and standard libraries. We can see that the DWT based algorithm is 50 times faster than the HHT based algorithm. For this reason, the DWT would be a better option in this particular case, where processing time matters. This does not mean that HHT could not be used at all in this context, though it will be very convenient to prioritize the method(s) with higher time efficiency. The reason is that this will give a valuable margin that could be used to address possible unexpected outcomes and perform other side tasks (e.g. adding significant flags to the data) before presenting the final results.

\begin{table}[htbp]
\begin{center}
\captionsetup{justification=centering , font=scriptsize, labelsep=newline}
\caption{\textsc{Processing time comparison, using a PC with Intel(R) Core(TM) processor i7-7500U CPU @ 2.70GHz, 2904MHz.}}

\begin{tabular}{c c }
\hline
 \parbox{5em}{Algorithm} & {Processing time (s)} \\ \hline
MA& 0.0009 \\ 
HHT& 0.12 \\ 
DWT& 0.0023 \\ \hline
\end{tabular}\vspace{0.2cm}
\label{table:comp}
\end{center}
\end{table}

\section{Conclusion}\label{conclusions}
The moving average filter is an efficient and appropriate solution for noise filtering when the target signals are affected by pure additive white Gaussian noise. Such a situation, however, is unlikely to take place for the ATS data recorded in the Mars environment. We propose the DWT and the HHT as the core for more powerful denoising methods, able to follow abrupt signal changes that may contain relevant side information. We have also proposed a procedure to validate the different methods' performance based on the comparison of the estimated noise standard deviation when using the controlled TVT data as a reference against the Martian ATS data, which is subject to a variety of external perturbations.  According to the results obtained, it may be correctly assessed that the proposed methods are effective in filtering the noise without potentially masking relevant signal features.
The HHT and DWT based filtering yield similar results in terms of denoising, but, taking into account the importance of the processing time for the mission, we propose the wavelet-based algorithm as the preferred choice because it is computationally more efficient.

It is to be noted that bringing environmental data from Mars to Earth for their processing and further study requires a tremendous effort, so that it may not make sense to radically smooth the ATS received data, thus removing some potential information content that could effectively cooperate with other sensors and measurements. This, in turn, can lead to a better understanding of the Mars environment. Further analysis of the REMS ATS data pre-processed with the techniques proposed here will surely ascertain the convenience of this approach.



%
\section*{Acknowledgment}
\addcontentsline{toc}{section}{Acknowledgment}
We want to thank the Editors and the Reviewers for all their valuable and constructive comments throughout the revision process, and for their positive review.

\bibliographystyle{IEEEtran}


%

%

\begin{IEEEbiographynophoto}
{Sof\'ia Zurita-Zurita} was born in Seville, Spain. She received the M.S  Degree in Telecommunications Engineering from the University of Seville, Spain, in 2008. She is currently  pursuing a Ph. D degree in the Centro de Astrobiolog\'ia in collaboration with the University of Alcala, both in Madrid, Spain. In parallel with her thesis, she is working at the Department of Instrumentation and Space Exploration in the  Centro de Astrobiolog\'ia, with TWINS and MEDA instruments, on board NASA's InSight and Mars 2020 missions.
\end{IEEEbiographynophoto}

\begin{IEEEbiographynophoto} 
{Francisco J. Escribano} (M'06, SM'16) received his degree in Telecommunications Engineering at ETSIT-UPM, Spain, and his Ph.D. degree at Universidad Rey Juan Carlos, Spain. He is currently Associate Professor at the Department of Signal Theory \& Communications of Universidad de Alcal\'{a}, Spain, where he is involved in several undergraduate and master courses in Telecommunications Engineering. He has been Visiting Researcher at the Politectnico di Torino, Italy, and at the \'{E}cole Polytechnique F\'{e}d\'{e}rale de Lausanne, Switzerland. His research activities and interests are focused on Signal Processing, Communication Systems at the PHY level and Information Theory. He has a number of publications in top level journals on the topics of index modulations for free-space optical communications, anytime reliable communication systems, and chaos-based coded modulations. He is currently working with Software-Defined Radio systems, and studying the application of advanced processing methods to the data recorded on Martian atmosphere.
\end{IEEEbiographynophoto}

\begin{IEEEbiographynophoto}
{Jos\'e S\'aez-Landete} was born in Valdeganga, Albacete, Spain, in 1977. He received the M.S. degree in physics from the Universidad de Zaragoza, Spain, in 2000, and the Ph.D. degree in physics from the Universidad Complutense de Madrid, in 2006. Since 2006, he has been with the Signal Theory and Communications Department of the Universidad de Alcal\'a, where he is currently an Associate Professor. His research interests include digital signal processing, digital communications, image processing, filter design, optimization and deep learning.
 \end{IEEEbiographynophoto}

\begin{IEEEbiographynophoto} 
{Jose A. Rodr\'{i}guez-Manfredi} (ORCID 0000-0003-0461-9815): Department of Instrumentation and Space Exploration, Centro de Astrobiologia - Instituto Nacional de T\'{e}cnica Aeroespacial, Madrid, Spain.
He received the Ph.D. degree in Engineering from the University of Seville (Spain) in 2007, within the Doctoral Program of Automation, Robotics and Telematics. He is the Principal Investigator of the TWINS instrument (onboard NASA's InSight mission) and the MEDA instrument (onboard NASA's Mars2020 / Perseverance mission to Mars). His current research interests include the development, testing, and deployment of advanced and innovative scientific instrumentation to explore and characterize other planets, moons, and extreme environments on Earth, from an Astrobiological perspective.
 \end{IEEEbiographynophoto}
%
%



\end{document}